\documentclass[a4paper]{article}
\usepackage{geometry}
\usepackage{authblk}
\usepackage{hyperref}
\usepackage{graphicx}
\usepackage{caption}
\usepackage{natbib}
\usepackage{multicol}
\usepackage{multirow}
\usepackage{booktabs}
\usepackage{aas_macros}
\newcommand\MyHead[2]{%
    \multicolumn{1}{l}{\parbox{#1}{\centering #2}}
}
\newcommand{\myimage}[1]{\begin{center}\includegraphics[angle=0, width=0.9\linewidth]{images/#1}\end{center}}

\newcommand{\myimageFour}[4]{\begin{center}\includegraphics[angle=0, width=0.45\linewidth]{images/#1}
		\includegraphics[angle=0, width=0.45\linewidth]{images/#2}\end{center}
	\begin{center}\includegraphics[angle=0, width=0.45\linewidth]{images/#3}
		\includegraphics[angle=0, width=0.45\linewidth]{images/#4}\end{center}}

\title{Isochrone fitting in the Gaia era. III. Distances, ages and masses from UniDAM using Gaia eDR3 data.}
\author[1]{A. Mints}
\affil[1]{Leibniz-Institut für Astrophysik Potsdam (AIP), Potsdam, Germany}
\begin{document}

\maketitle

\begin{abstract}
We present estimates of distances, ages and masses for over  million stars. These estimates are derived from 
the combination of spectrophotometric data and Gaia eDR3 parallaxes.
For that, we used the previously published Unified tool to estimate Distances, Ages, and Masses (UniDAM).
\end{abstract}

\section{Introduction}
In \citep{2017A&A...604A.108M} we presented the Unified tool to estimate Distances, Ages, and Masses (UniDAM\footnote{UniDAM source code is available at \url{https://github.com/minzastro/unidam}.}). UniDAM uses Bayesian scheme to derive distances, ages and masses for stars from spectrophotometric data.
This tool was further updated (see \citep{2018arXiv180406578M}), to allow for the use of parallax data in isochrone fitting. Once Gaia DR2 was released, \citep{2016A&A...595A...1G,2018arXiv180409365G} we produced 
a new catalogue of distance, ages and masses for over 3,5 million stars \citep{2018arXiv180501640M}. 
The recently released version 2.3 of UniDAM also allows the use of asteroseismic data.

Several spectroscopic surveys have made new releases recently: RAVE \citep[final DR6][]{2020AJ....160...83S},
LAMOST DR6 \citep[also containing the first data from medium-resolution survey]{2015RAA....15.1095L},
GALAH+ DR3 \citep{2020arXiv201102505B} and APOGEE DR16 \citep{2020ApJS..249....3A}.

\begin{figure}
	\myimage{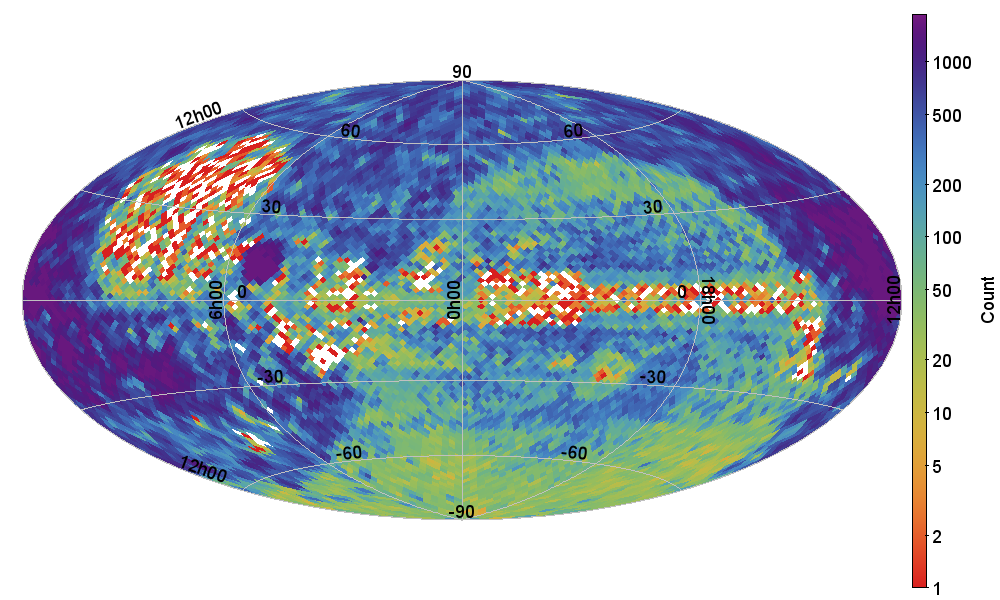}
	\caption{Distribution in galactic coordinates of unique sources in the published catalogue.}\label{fig:sky}
\end{figure}

When supplemented by other surveys already processed in \citep{2018arXiv180501640M}, this gives almost 8 millions of measurements for almost 6 million sources. They are distributed over almost entire sky (see \autoref{fig:sky}), with some gaps in southern Galactic disc and around northern equatorial pole. An overdensity at about $l = 75^o$ and $b = +13^o$ corresponds to the \textit{Kepler} field of view, that was extensively studied by APOGEE and LAMOST \citep[see, for example][]{LamostKepler, APOKASC}. 

All surveys used in the current work are listed in \autoref{tbl:overlap}.

\begin{table*}[t]
    \centering
            \begin{tabular}{l r r p{5cm}} \toprule
    Survey              & \MyHead{2cm}{Input catalogue size}     & 
    \MyHead{3cm}{Stars with estimates done using Gaia eDR3 parallaxes}  & \MyHead{2cm}{Reference}  \\ \midrule
    APOGEE (DR16)       & 473,307   & 326,884   & \citet{2020ApJS..249....3A} \\
    Bensby              & 714       & 547        & \citet{Bensby2014} \\
    Gaia-ESO (DR3)      & 25,533    & 20,127     & \citet{2012Msngr.147...25G}, \citet{GaiaESO_dr3} \\
    GALAH (DR3)         & 564,620   & 505,403   & \citet{2020arXiv201102505B} \\
    GCS                 & 13,565    & 7,633     & \citet{GCS} \\
    LAMOST (DR6)        & 5,581,266 & 4,377,103 & \citet{2015RAA....15.1095L} \\
    LAMOST MRS (DR6)    & 328,187   & 223,407  & \citet{2015RAA....15.1095L} \\	
    RAVE (DR6)          & 491,349   & 347,211   & \citet{2020AJ....160...83S} \\
    SEGUE               & 235,595   & 180,012   & \citet{2009AJ....137.4377Y} \\
    Total (unique sources) & 5,856,273 & 4,616,931 & \\ \bottomrule& 
    \end{tabular}
    \caption{Total number of sources with 2MASS/AllWISE photometry, 
    	and Gaia eDR3 overlap for different surveys.}\label{tbl:overlap}
\end{table*}

\section{Gaia eDR3 data}
Gaia early data release 3 \citep[Gaia eDR3,][]{GaiaEDR3} was released on December 3rd 2020. It contains among other data parallaxes and proper motions for around 1.468 billion sources. As compared to the previous Gaia DR2, precision and accuracy of astrometry was substantially improved as additional 12 month more of observations were used and as calibration was largely improved.
We have crossmatched all sources in our catalogues listed in \autoref{tbl:overlap} with Gaia eDR3 data using Vizier XMatch tool \footnote{\url{http://cdsxmatch.u-strasbg.fr}}.

As stated in \citet{GaiaEDR3}, Gaia eDR3 parallaxes seem to have a systematic offset of $-17 \mu$as -- about a factor of 3 lower than in Gaia DR2. This offset was added to parallax values used in UniDAM processing.

\section{Results and discussion}
\begin{figure}
	 \myimage{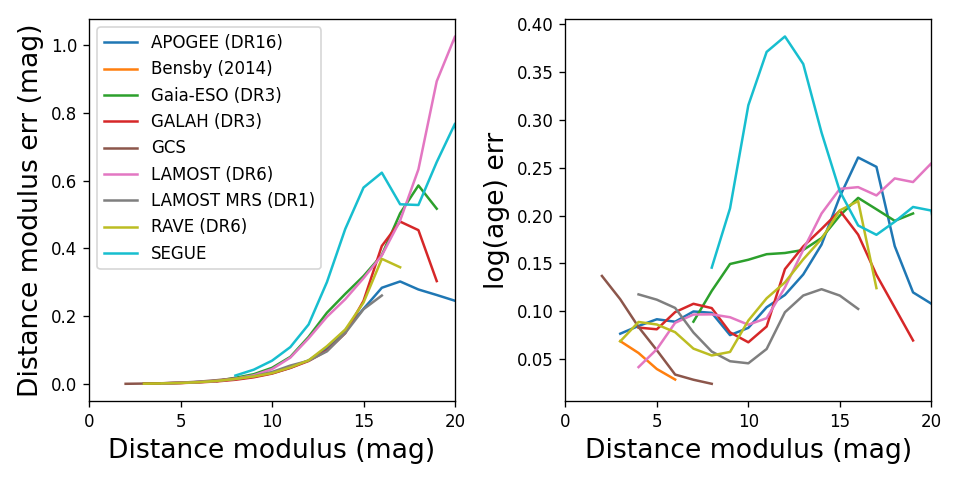}
	 \caption{Uncertainties of UniDAM results with Gaia eDR3 parallaxes in distance modulus (left) and log(age) (right) for different spectroscopic surveys.}\label{fig:results}
\end{figure}

In Figure \ref{fig:results} we show how median uncertainties of distance modulus and log(age) derived by UniDAM with the use of Gaia eDR3 data compare with results obtained with Gaia DR2 parallaxes.

Most surveys show very similar distance modulus uncertainty behaviour, as distance modulus uncertainty depends primarily on the parallax uncertainty. There are some clear exceptions though. SEGUE, LAMOST and Gaia-ESO surveys show larger uncertainties, when compared to other surveys. This is because these surveys target larger fraction of main sequence stars that are typically much fainter at the same distance. Those fainter stars have systematically larger Gaia eDR3 parallax uncertainties and hence larger derived distance modulus uncertainties. On the other side, APOGEE stars have much lower distance modulus uncertainty at $\mu_d > 15^m$. At this distances, a smaller uncertainty in surface gravity that APOGEE stars have starts to play a role. With Gaia eDR3 parallaxes becoming more uncertain for distant stars, UniDAM constrains distance mainly from spectrophotometric data -- and thus having lower uncertainties in surface gravity for APOGEE provides lower uncertainty in distance modulus.

Uncertainty of log(age) shown in the right panel of \autoref{fig:results} depends not only on the uncertainties in spectrophotometric parameters and parallaxes, but also on the positions of stars in the Hertzsprung-Russell diagram: for stars on the main sequence age is much less constrained by spectrophotometric data. Hence the larger is the fraction of main-sequence stars in the given distance bin -- the larger the median uncertainty in log(age). As distance modulus increases, the fraction of main sequence stars decreases, and with it decreases the median uncertainty in log(age). With distance modulus increasing beyond $\mu_d \approx 10^m$, median uncertainty in log(age) starts to increase, due to uncertainties of spectrophotometric data increasing with distance. This upturn happens at larger distances ($\mu_d \approx 15^m$) for SEGUE survey, as its main-sequence sample extends to larger distances.

\begin{figure}
	\myimageFour{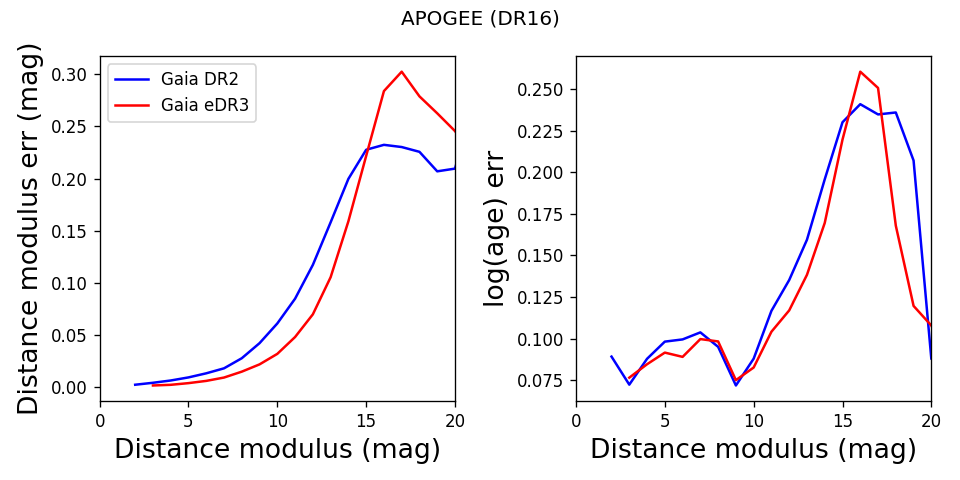}{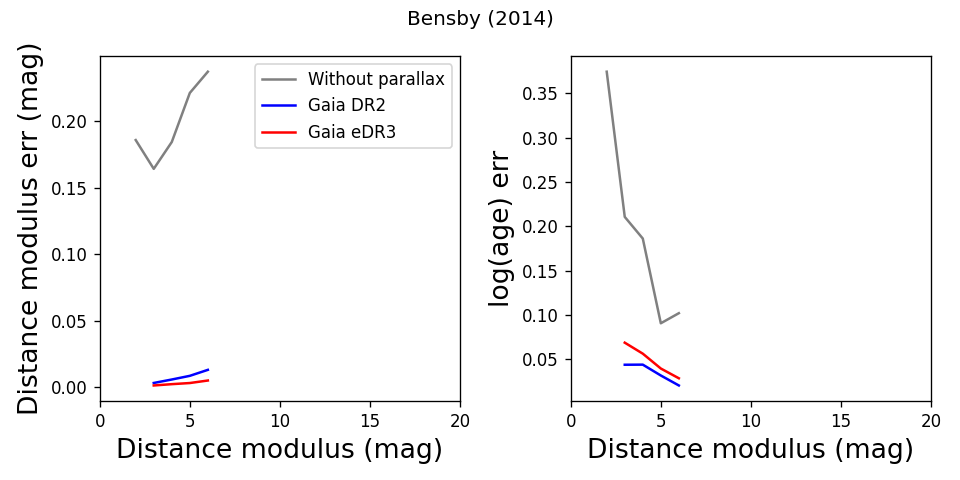}{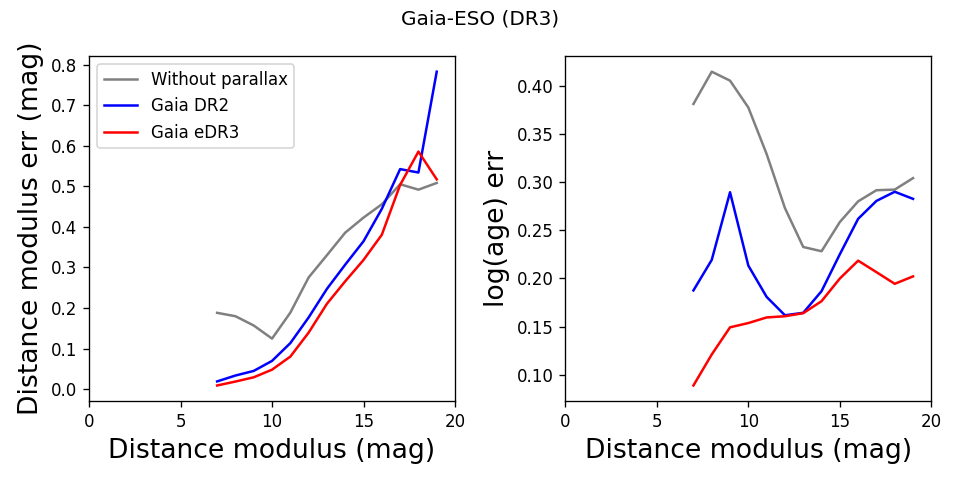}{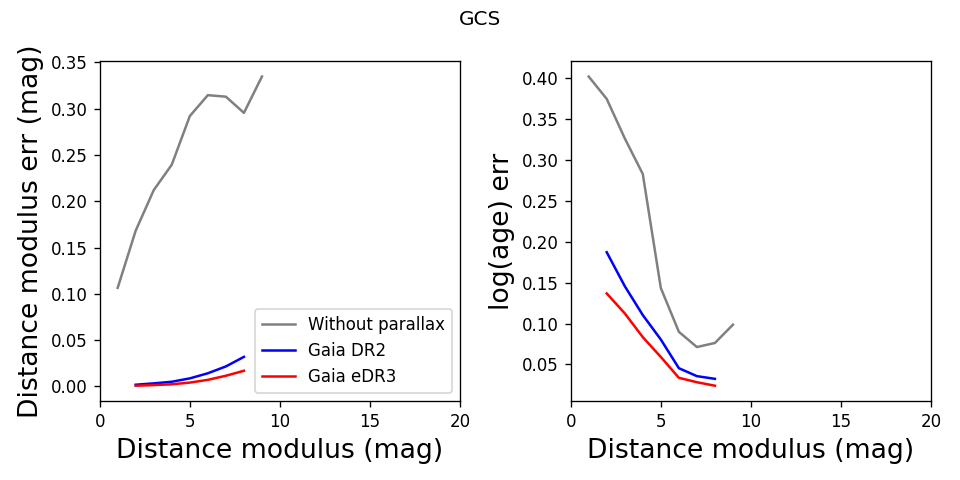}
	\myimageFour{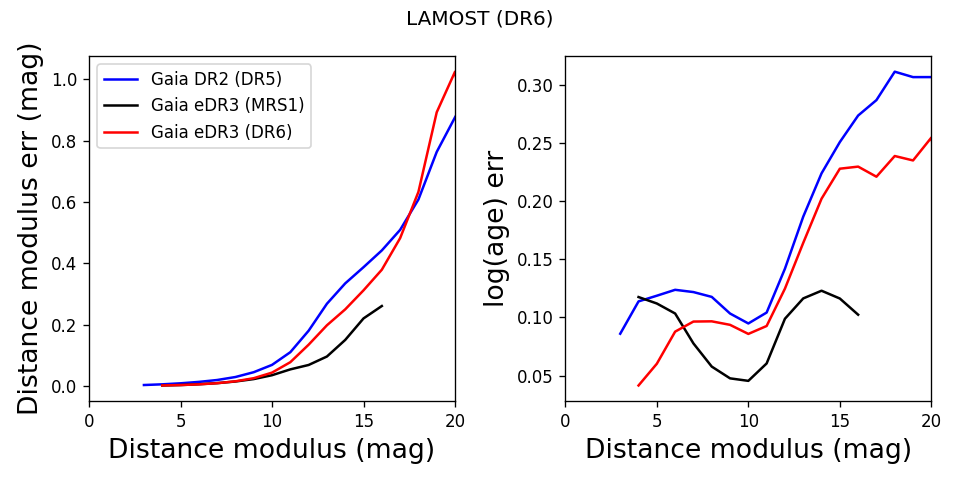}{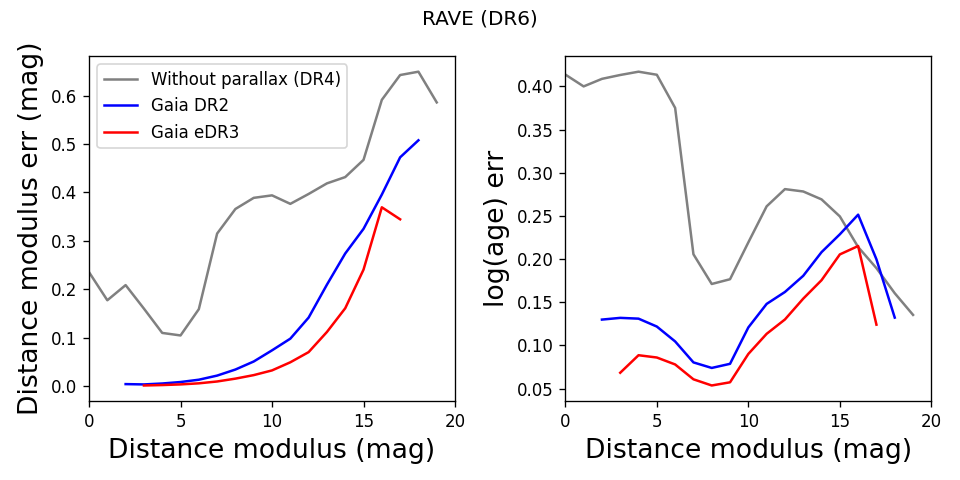}{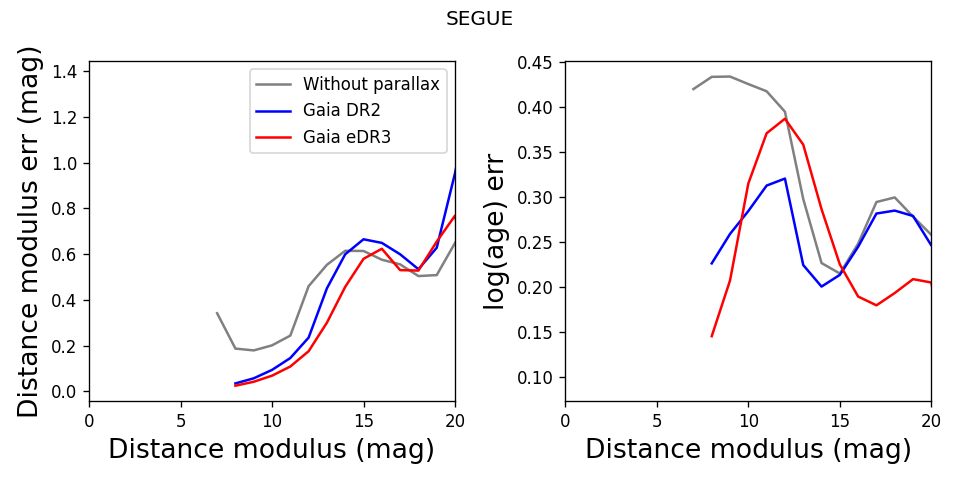}{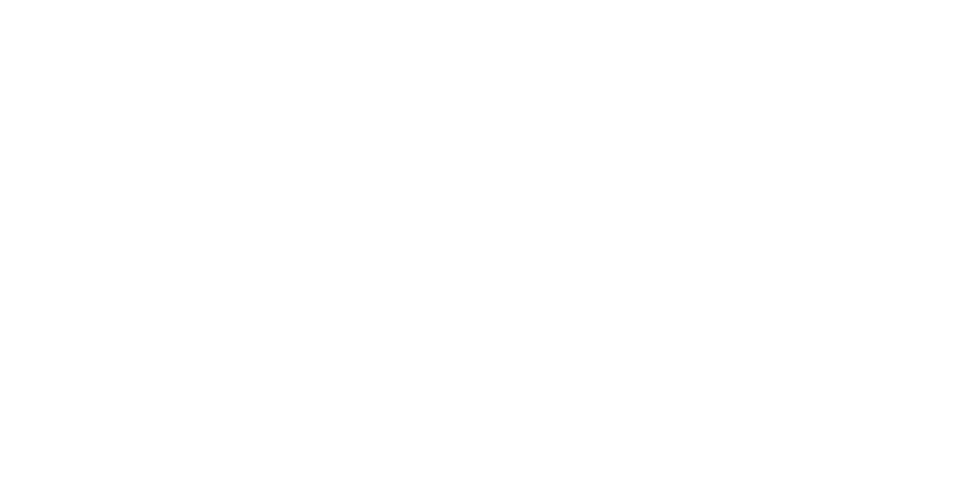}
	\caption{Uncertainties of UniDAM results in distance modulus and log(age)s for several surveys. Plots show data without parallaxes (where available, grey line), with Gaia DR2 parallaxes (blue line) and with Gaia eDR3 parallaxes (red line).}\label{fig:persurvey}
\end{figure}

In \autoref{fig:persurvey} we compare uncertainties in distance modulus and log(age) without parallax data (where available), with Gaia DR2 data and Gaia eDR3 data to show the improvement we get from the newest Gaia eDR3.
It is clear that Gaia eDR3 parallaxes allow to decrease uncertainties.

We publish results of UniDAM with Gaia eDR3 parallaxes on \citet{alexey_mints_2020_4312713}. An important feature of UniDAM is that it can produce more than one solution for a given star. This allows to separate out solutions for different evolutionary stages or, more generally, for cases when posterior distributions in distance modulus or log(age) are multi-modal. Values reported for each solution allow detailed reconstruction of posterior distribution in every parameter (see UniDAM homepage\footnote{\url{https://github.com/minzastro/unidam}} for details).
In total, our data contains 5988327 solutions for 4616931 unique stars. This is about 20 percent less than the size of the input catalogue, because for many stars no isochrone fit is possible. There may be several reasons for that, for example:
\begin{itemize}
	\item Incomplete spectroscopic data in the survey.
	\item There are no 2MASS/AllWISE photometry for the star -- it is either too faint, too bright or suffer from some photometric issues. 
	\item No Gaia eDR3 parallax available -- it is either too faint, too bright or suffer from some astrometric issues. 
	\item There is no model that fits spectroscopic parameters. This can happen if the star is outside of PARSEC model range (it spans, for example, only metallicities between -2.18 and +0.5 dex). Untypical and non-stellar spectra can also produce a combination of spectroscopic parameters that has no match in PARSEC models -- for such objects no solution is found too.
	\item There are inconsistencies in the input data. Those can arise for example when the observed object is a binary or a blend of two or more stars (and thus is brighter than expected from spectroscopic parameters and parallax). Inconsistency can also be a sign of either an outlier in the input data or erroneous match between spectroscopic survey and 2MASS, AllWISE and Gaia catalogues.
\end{itemize}

\section*{Acknowledgements}
\tiny{
This work has made use of data from the European Space Agency (ESA) mission
\textit{Gaia} (\url{https://www.cosmos.esa.int/gaia}), processed by the {\textit Gaia}
Data Processing and Analysis Consortium (DPAC,
\url{https://www.cosmos.esa.int/web/gaia/dpac/consortium}). Funding for the DPAC
has been provided by national institutions, in particular the institutions
participating in the {\it Gaia} Multilateral Agreement.

Funding for Rave has been provided by: the Leibniz Institute for Astrophysics Potsdam (AIP); the Australian Astronomical Observatory; the Australian National University; the Australian Research Council; the French National Research Agency; the German Research Foundation (SPP 1177 and SFB 881); the European Research Council (ERC-StG 240271 Galactica); the Istituto Nazionale di Astrofisica at Padova; The Johns Hopkins University; the National Science Foundation of the USA (AST-0908326); the W. M. Keck foundation; the Macquarie University; the Netherlands Research School for Astronomy; the Natural Sciences and Engineering Research Council of Canada; the Slovenian Research Agency; the Swiss National Science Foundation; the Science \& Technology FacilitiesCouncil of the UK; Opticon; Strasbourg Observatory; and the Universities of Basel, Groningen, Heidelberg and Sydney.

Guoshoujing Telescope (the Large Sky Area Multi-Object Fiber Spectroscopic Telescope LAMOST) is a National Major Scientific Project built by the Chinese Academy of Sciences. Funding for the project has been provided by the National Development and Reform Commission. LAMOST is operated and managed by the National Astronomical Observatories, Chinese Academy of Sciences.

Based on data acquired through the Australian Astronomical Observatory, under programs: A/2013B/13 (The GALAH pilot survey);
A/2014A/25, A/2015A/19, A2017A/18 (The GALAH survey phase
1), A2018 A/18 (Open clusters with HERMES), A2019A/1 (Hierarchical star formation in Ori OB1), A2019A/15 (The GALAH survey phase 2), A/2015B/19, A/2016A/22, A/2016B/10, A/2017B/16,
A/2018B/15 (The HERMES-TESS program), and A/2015A/3,
A/2015B/1, A/2015B/19, A/2016A/22, A/2016B/12, A/2017A/14,
(The HERMES K2-follow-up program). We acknowledge the traditional owners of the land on which the AAT stands, the Gamilaraay
people, and pay our respects to elders past and present.

This research has made use of the VizieR catalogue access tool, CDS, Strasbourg, France. 
This research made use of the cross-match service provided by CDS, Strasbourg.
This research made use of SciPy \citep{Virtanen_2020}.
This research made use of Astropy, a community-developed core Python package for Astronomy \citep{2018AJ....156..123A, 2013A&A...558A..33A}.
Based on data products from observations made with ESO Telescopes at the La Silla Paranal Observatory under programme ID 188.B-3002. These data products have been processed by the Cam-bridge Astronomy Survey Unit (CASU) at the Institute of Astronomy, University of Cambridge, and by the FLAMES/UVES reduction team at INAF/Osservatorio Astrofisico di Arcetri. These data have been obtained from the Gaia-ESO Survey Data Archive, prepared and hosted by the Wide Field Astronomy Unit, Institute for Astronomy, University of Edinburgh, which is funded by the UK Science and Technology Facilities Council. 
This publication makes use of data products from the Wide-field Infrared Survey Explorer, which is a joint project of the University of California, Los Angeles, and the Jet Propulsion Laboratory/California Institute of Technology, and NEOWISE, which is a project of the Jet Propulsion Laboratory/California Institute of Technology. WISE and NEOWISE are funded by the National Aeronautics and Space Administration 
This publication makes use of data products from the Two Micron All Sky Survey, which is a joint project of the University of Massachusetts and the Infrared Processing and Analysis Center/California Institute of Technology, funded by the National Aeronautics and Space Administration and the National Science Foundation. 

}

\bibliographystyle{aa}
\bibliography{bibliography}

\end{document}